\documentclass[preprint,amsmath,showpacs]{revtex4}
\begin{document}
\title{THE DMPK EQUATION FOR MESOSCOPIC QUANTUM TRANSPORT REVISITED}
\author{Jean Heinrichs}
\email{J.Heinrichs@ulg.ac.be} \affiliation{D\'{e}partement de
Physique, B{5a}, Universit\'{e} de Li\`{e}ge, Sart Tilman, B-{4000}
Li\`{e}ge, Belgium}
\date{\today}
\begin{abstract}
A recent Drude model description of the metallic regime and of a channel- averaged elastic mean free path (mfp), $\ell_0$, in an $N$-channel tight-binding wire identifies the Thouless localization length, $N\ell_0$, as a proper lower bound of macroscopic length scales ("mean free path") for the DMPK equation describing the localized regime of the wire. The mfp $\ell_0$ leads to a metallic regime which is consistent with Dorokhov's microscopic transmission analysis in terms of a nominal elastic mfp. On the other hand, the validity of Mello's derivation of universal conductance fluctuations in the metallic regime based on the DMPK equation is restored if the mfp $\ell'$, of order $N\ell_0$, in that equation is replaced by the correct mean free path $\ell_0$.

\end{abstract}
\maketitle
The DMPK equation is unquestionably an important tool for studying quantum transport in disordered multichannel wires \cite{1,2,3,4,5,6,7,8,9,10,11}, as shown, in particular, by the extensive reviews \cite{9,10} and monographs \cite{8,11} which discuss it along with applications \cite{12}.
However, an unsatisfactory feature of the DMPK equation has emerged recently through the comparison of exact analytic results for localization lengths in few-channel disordered wires ($N=2,3$) \cite{13,14,15} with the Thouless localization length \cite{16},
\begin{equation}\label{eq1}
\xi\sim N\ell\quad ,
\end{equation}
using the mean free path (mfp) defined in the DMPK approach \cite{2,3,4,5,8,9,10,11}. The latter may be expressed as \cite{10,11}
\begin{equation}\label{eq2}
\frac{1}{\ell}=
\frac{1}{NL}\sum^{N}_{i,j=1}
\langle |r_{ij}^{(N)}|^2\rangle\quad ,
\end{equation}
where the $r_{ij}^{(N)}$ denote the elements of the $N\times N$ reflection amplitude submatrix $\hat r^{(N)}$ of the scattering matrix ($\widehat S$)
of the quasi-one-dimensional wire of length $L$ \cite{13} and $\langle\ldots\rangle$ means averaging over the disorder. We note that the factor $N^{-1}$ in this expression is arbitrary and has been introduced for convenience \cite{2} without further justification.

In Refs \cite{13,14,15} we presented exact, analytical calculations (for weak disorder) of reflection and transmission matrix elements for two- and three-channel systems (for open- as well as for periodic boundary conditions) and also for Dorokhov's equivalent-channel model \cite{15}. In \cite{13} and \cite{15} we restricted to systems where all modes at the fermi level are propagating while in \cite{14} we considered the case where propagating and evanescent modes coexist at the fermi level.
Using \eqref{eq2}, we obtained in all cases considered that $\xi/\ell=2$, which would indicate the absence of a metallic regime, as defined for length scales $\ell\ll L\ll\xi$, in quasi-one-dimensional systems. This had led us to derive a precise physical model for the mean free path in a multi-channel tight-binding wire \cite{17} based on the Drude model of metallic conduction \cite{18} and the use of Ohm's law. Our model readily leads to the Thouless formula \eqref{eq1} in terms of a channel-averaged mean free path \cite{19} given by
\begin{equation}\label{eq3}
\frac{1}{\ell_0}=
\frac{1}{2L}
\text{Trace }\langle \hat r^{(N)}\hat r^{(N)+}\rangle\quad ,
\end{equation}
where $\hat r^{(N)}$ is the $N\times N$ reflection amplitude sub-matrix of the $\widehat S$-matrix. It follows from \eqref{eq2} and \eqref{eq3} that
\begin{equation}\label{eq4}
\frac{\ell_0}{\ell}=\frac{2}{N}
\quad ,
\end{equation}
which demonstrates, in particular, the irrelevance of the averaging over incoming channels \cite{20} which is included in \eqref{eq2}.  We also recall that \eqref{eq3} has been obtained in the well-defined weak disorder limit where the average reflection coefficient per channel is close to zero and the average transmission coefficient is close to one.

Finally, we note that our results in \cite{13,14} for -two- and three-channel tight-binding systems have recently been generalized by Gasparian and Suzuki \cite{21} in the case of quasi-one-dimensional systems with an arbitrary number of channels, using an ingeneous determinant appproach to the $S$-matrix elements \cite{22} and Eq.\eqref{eq3} for defining the mean free path.
These results lead to the Thouless localization length $\xi\sim N\ell_0$ for all cases analyzed in \cite{13,14,15,21}, which implies the existence of well-defined metallic domains for $\ell_0\ll L\ll\xi$.

We now observe that the results of \cite{13,14,15,21}, with the definition \eqref{eq3} of the mean free-path are consistent with Dorokhov's earlier probabilistic transmission coefficients analysis in [1a] for a similarly defined coupled tight-binding disordered channels model. In particular, Dorokhov [1a] obtains that for $L\gg N\ell$ (where $\ell$ stands for the nominal elastic mean free path in [1a]) the evolution equation for the probability distribution of transmission coefficients in the $N$ coupled channels system reduces to a set of equations describing localization in the individual channels with $N$ distinct localization lengths up to a maximum length $\xi=N\ell$, namely the Thouless localization length (see Eq.(14) of Ref.[1a]). This agrees indeed with our earlier results for $\ell$ defined by $\ell_0$ in \eqref{eq3} and establishes the primordial role of the Dorokhov probabilistic analysis of the Anderson model for demonstrating the metallic regime of the wire, which is not obtained with the mean free path definition used in the DMPK equation \cite{17}.

The description of the metallic regime by the elastic mfp \eqref{eq3} and the Thouless length $\xi\sim N\ell_0$ has an immediate consequence for the analysis of the localized regime by means of the macroscopic DMPK equation for the statistical distribution of the transfer matrix \cite{2,9}.  Indeed, for $N>>1$ the Thouless length bounding the metallic regime acts as a typical lower limit of macroscopic length scales for the validity of the DMPK equation for describing the localized regime, which sets in for $L>N\ell_0$.  

In terms of the mean free path \eqref{eq3} the DMPK given in \cite{9} now reads

\begin{subequations}
\begin{align}
\ell_0\frac{\partial P}{\partial L}
&=\frac{2}{\beta N+2-\beta}
\sum^N_{n=1}
\frac{\partial}{\partial\lambda_n}
\lambda_n (1+\lambda_n)
J\frac{\partial}{\partial\lambda_n}\frac{P}{J}\label{eq5a}\quad ,\\ 
J
&=\prod^N_{i=1}\prod^N_{j=i-1}
|\lambda_j-\lambda_i|^\beta,
\lambda_i=(1-T_i)/T_i\label{eq5b}\quad ,
L>N\ell_0
\quad .
\end{align}
\end{subequations}
Here $T_1,T_2,\ldots T_N$ denote the eigenvalues of transmission matrices, $\hat t\hat t^+$, $\hat{t'}\hat{t'}^+$ (from left to right and from right to left, respectively) or of corresponding reflection matrices $\hat r\hat r^+$, $\hat{r'}\hat{r'}^+$. The parameter $\beta$ is the ensemble symmetry parameter which depends on the presence or not of time reversal- and/or not of spin-rotation symmetry \cite{8,9}. The DMPK equation (\ref{eq5a},\ref{eq5b}) describes the evolution of the distribution $P(\lambda_1,\lambda_2,\ldots\lambda_n)$ of the variables $\lambda_1,\lambda_2,\ldots\lambda_n$ as a function of length $L$.

We conclude by discussing two important consequences of the introduction of the proper elastic mfp of Eq.(\ref{eq3}) in the DMPK equation (\ref{eq5a},\ref{eq5b}) which is compatible with the existence of a diffusive metallic regime for lengths 

\begin{equation}\label{eq6}
\ell_0<<L\leq N\ell_0 
\end{equation}
(with $N\ell_0$ viewed as a macroscopic length for $N>>1$) and leads to an insulating regime for larger lengths.  The first important consequence is the existence of universal conductance fluctuations in the metallic regime \eqref{eq6} in multichannel wires, which have been demonstrated by Mello \cite{3} by solving the DMPK equation for lengths $L$ much larger than a nominal mfp $\ell'$ and $N>>1$.  These results followed the earlier derivation of ucf in the more general cases of two- and three-dimensional systems by Lee and Stone \cite{23}, using diagrammatic perturbation theory.  Now, the nominal mfp $\ell'$ in Mello's analysis \cite{3} could be identified with $\ell_0$ in \eqref{eq3} but not with the DMPK mfp $\ell$ in \eqref{eq2}, of the order of the Thouless length which borders the metallic regime\cite{17}.  The latter clearly precludes Mello's expansion of the DMPK equation for large $L/\ell$ in the metallic regime.  In conclusion, the use of the mfp \eqref{eq3} in the DMPK equation is primordial for validating Mello's derivation of ucf in quasi-one-dimensional systems.  It may be useful as well for future numerical applications of the DMPK equation, in the metallic regime where it appears to remain valid provided the DMPK mfp \eqref{eq2} is replaced by the mfp \eqref{eq3}.

Another important result obtained from (\ref{eq5a},\ref{eq5b}) is the detailed form of the localization length in the insulating domain for $L>N\ell_0$ derived by Beenakker \cite{9}, which now reads
\begin{equation}\label{eq7}
\xi=(\beta N+2-\beta)\ell_0
\quad .
\end{equation}
In the presence of time-reversal symmetry ($\beta=1$) this expression is remarkably close to the Thouless form $\xi\sim N\ell_0$.


\begin{thebibliography}{unsrt}
\bibitem{12} In the context of this Communication it is important to note the fundamental difference between the approaches of Dorokhov \cite{1} and of Mello, Pereyra and Kumar \cite{2} for studying transport in multichannel disordered conductors. The approach of Dorokhov is microscopic, using the Anderson one-orbital tight-binding model. On the other hand, the analysis of Mello, Pereyra and Kumar uses a macroscopic maximum entropy model with no direct reference to microscopic aspects. In spite of this the acronym DMPK has been generally adopted in the literature, for referring to the equation derived from the phenomenological macroscopic approach.
\bibitem{19} The need for defining an average channel mean free path arises from the fact that the individual channels of the wire are associated with a collection of discrete fermi points on a "fixed line surface" in momentum space.
\bibitem{20} The definition of the mfp used in the DMPK equation has been rationalized in \cite{4,5,11} as describing the reflection coefficient of the wire summed over outgoing channels and averaged over incoming ones.  However, the pertinence of this process is not supported by our subsequent derivation of the elastic mfp in a quasi-one-dimensional wire \cite{17} recalled above (see Eq. \eqref{eq4}).
\end{thebibliography}
\end{document}